# Protein Enrichment by Flotation: Experiment and Modeling

Behnam Keshavarzi[1,2,*], Thomas Krause[3], Sidra Sikandar[1], Karin Schwarzenberger [1,2], Kerstin Eckert[1,2], Marion B. Ansorge-Schumacher[3] and Sascha Heitkam [1,2,*]

[1] Institute of Process Engineering and Environmental Technology, TU Dresden, 01062 Dresden, Germany

[2] Institute of Fluid Dynamics, Helmholtz-Zentrum Dresden-Rossendorf, Bautzner Landstrasse 400, 01328 Dresden, Germany

[3] Department of Molecular Biotechnology, TU Dresden, 01062 Dresden, Germany

**Graphical Abstract:**

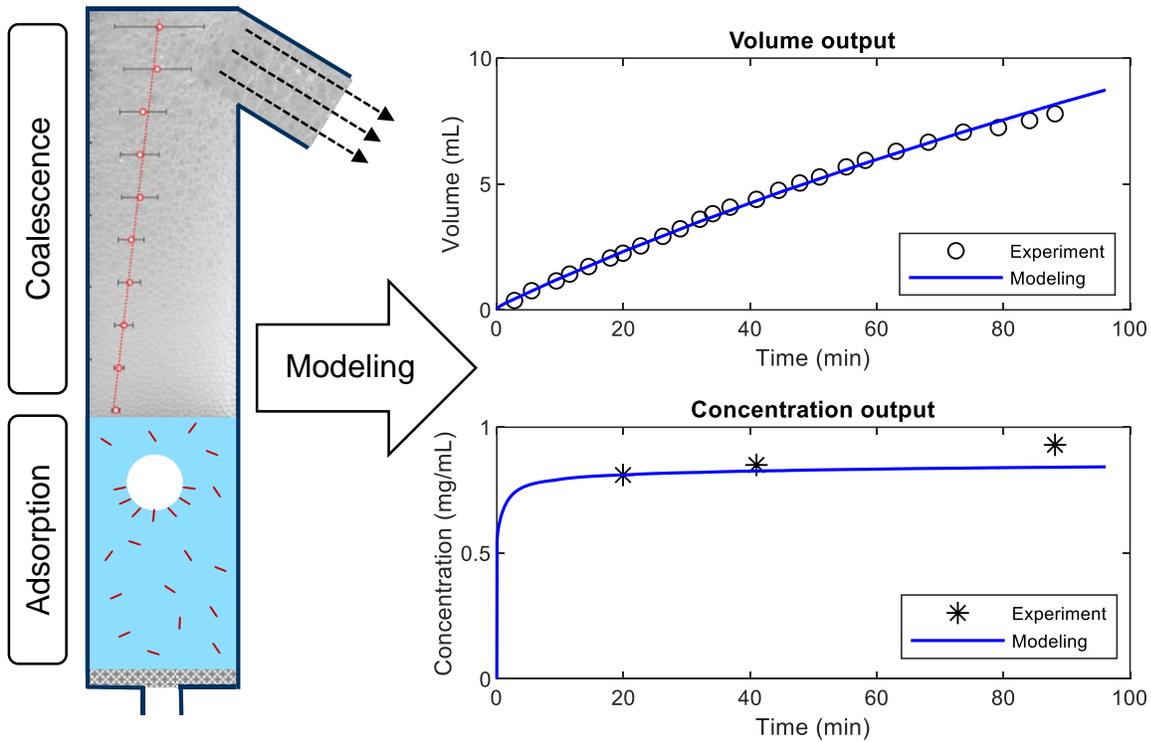

**Abstract**

Protein flotation is a process in which protein molecules are enriched by adsorption at rising bubbles. The bubbles then form a foam above the solution, where the liquid drains down and the dried foam, which is concentrated in protein, is extracted. Here, the recovery rate and purity of the extract depend on foam stability, surface coverage, bubble size, gas flow rate, etc.

---

[*] Corresponding author: Institute of Process Engineering and Environmental Technology, TU Dresden, 01062 Dresden, Germany. Email addresses: sascha.heitkam@tu-dresden.de, behnam.keshavarzi@tu-dresden.de.

In this work, we performed flotation experiments using bovine serum albumin (BSA). In addition, an unsteady state simulation of the protein flotation process was carried out by numerically solving the liquid drainage equation in the foam. Thereby, the extracted liquid volume and protein concentration at the outlet were calculated with time. Required quantities such as foam stability, interface coverage or bubble size distribution were measured in corresponding experiments and were fed into the model.

The experiments showed that the foam coalescence accelerates the liquid drainage leading to dryer extract and higher protein enrichment. The modeling also reproduced the liquid recovery and extract concentration of the flotation tests within a reasonable error range. The modeling solely relies on experimental inputs and does not require any tuning parameters. It can be further used for optimization or up-scaling of protein flotation.

# 1 Introduction

In flotation, the target materials adsorb on the surface of the rising bubbles based on their hydrophobic properties and are collected in the foam. The liquid drains down through the foam column while the dried foam flows over the weir, yielding the target material in high concentration. Flotation has been implemented in many fields such as mineral processing [1], wastewater treatment [2] as well as protein separation [3-10], which is the subject of this work. In protein flotation, the protein molecules adsorb on the rising bubbles via their hydrophobic sites and are carried upward with the bubbles. One of the main characteristics of protein flotation is that no surfactant is used and the protein itself provides the foamability, as surfactants influence the protein structure [11-12]. Moreover, surfactants might be costly, and the subsequent separation of proteins from surfactants is difficult. Since the proteins have to ensure foamability, understanding their adsorption on the water-air interface and their behavior in the foam is essential. Together with an adequate description of the transport processes in the liquid and foam phase, this provides a tool to simulate the protein flotation process, which has been lacking until now.

Hence, in this work, we firstly utilize profile analysis tensiometry (PAT) to characterize the interfacial properties of the protein under quiescent conditions. A rough estimation of the protein adsorption on the rising bubbles can be obtained through surface tension (ST) measurements under similar flow conditions. Therefore, we emulate the bubble rise by performing flow-on-bubble experiments coupled with PAT, as reported earlier [13]. Now, the unsteady transport of liquid in the rising foam is considered, relying on bubble size and superficial gas flow rate. The bubble size in the liquid column is evaluated by image analysis of the rising bubbles. For the foam column, a linear change of the bubble size versus foam height due to coalescence is considered, whose slope is also obtained from image analysis. The liquid permeability through the foam column is calculated using the Carman-Kozeny equation [14] with an estimation for the specific area by Pitios et al. for a wide liquid content range [15]. Altogether, the modeling yields the liquid recovery and protein concentration in the extract as key parameters for the flotation process. The modeling is finally validated with experimental results. This further provides the basis to control and optimize protein flotation in technological applications.

## 2   Experimental setup and procedure

### 2.1   Materials

The bovine serum albumin (BSA, 98 % purity, Sigma Aldrich, Germany) solutions were prepared in a buffer solution with pH $7.3 \pm 0.1$. For each set of experiments, a main stock solution was made by adding a pre-weighted amount of BSA powder to the buffer solution and gently stirring for 15 minutes using a magnetic stirrer. The lower concentrations were diluted from the stock solution with further buffer addition. All protein solutions were stored at a temperature of 4 °C ± 1 and were used within 24 hours after preparation of the main stock solution. For the buffer solution, 12.81 g/l potassium hydrogenphosphate ($K_2HPO_4$) and 3.60 g/l of potassium dihydrogenphosphate ($KH_2PO_4$), (both Sigma Aldrich, Germany), were dissolved in deionized water.

## 2.2 Surface tension tests

We used profile analysis tensiometry (PAT; Sinterface, Germany) for dynamic ST measurements of the solutions [16]. In this work, we employed the buoyant bubble method. For each experiment, the BSA solution was firstly poured in the cuvette. Then, a bubble was generated by controlled injection of air through a capillary tube with 1 mm outer diameter until the bubble reached the desired volume. Afterwards, the ST was recorded with time.

## 2.3 Flow-on-bubble tests

The influence of convective mass transfer on the adsorption of BSA on the surface of the rising bubble in the flotation column is evaluated by flow-on-bubble tests. A nozzle with an inner diameter of 2.5 mm and an outflow line were added to the cell in the PAT device to apply flow directly on the bubble. When the flow is applied on the bubble, the bubble loses its Laplacian shape and a correct ST measurement through PAT is not possible. Therefore, the ST could be recorded before and after the flow. The schematic of the flow-on-bubble setup is shown in Figure 1-B together with a graph describing the ST changes during the test (Figure 1-C). During the flow phase, a balance between flow-induced shear stress and Marangoni stress results in an inhomogeneous distribution of the adsorbed material along the bubble surface. However, when the flow is stopped, the Marangoni stress again spreads the adsorbed material evenly along the surface [17].

A Harvard reciprocating pump consisting of two individual pumps was used to inject the fluid through the nozzle towards the bubble and to withdraw the same amount of liquid from the cell, keeping a constant liquid volume in the cuvette. For each test, the cuvette and the liquid supply line from the syringe was filled with the desired solution. Then, a bubble was formed at the tip of the capillary tube in the cuvette. The inflow on the bubble from the nozzle (located approximately 2 mm above the bubble top) was initiated 3 seconds after the bubble reached its final volume of 8 $mm^3$. When the flow was stopped, a waiting time of 2 seconds was kept to let the flow cease and then the ST was recorded as the final ST of the bubble. Each test was repeated at least three times and an average value was reported.

In our tests, the nozzle inner diameter is in the same order of magnitude as the bubble diameter ($\cong$ 2.4 mm). To improve this model experiment, a considerably larger nozzle diameter could be applied to assure a more homogeneous flow field around the bubble. However, the main purpose here is to induce convection to mimic the mass transfer mechanism at the rising bubble surface. Therefore, our approximation with an inhomogeneous flow field is acceptable because in the real situation of a rising bubble, mass transfer mainly takes place at the top of the bubble while the adsorbed material accumulates at the bottom and hinders further adsorption [18].

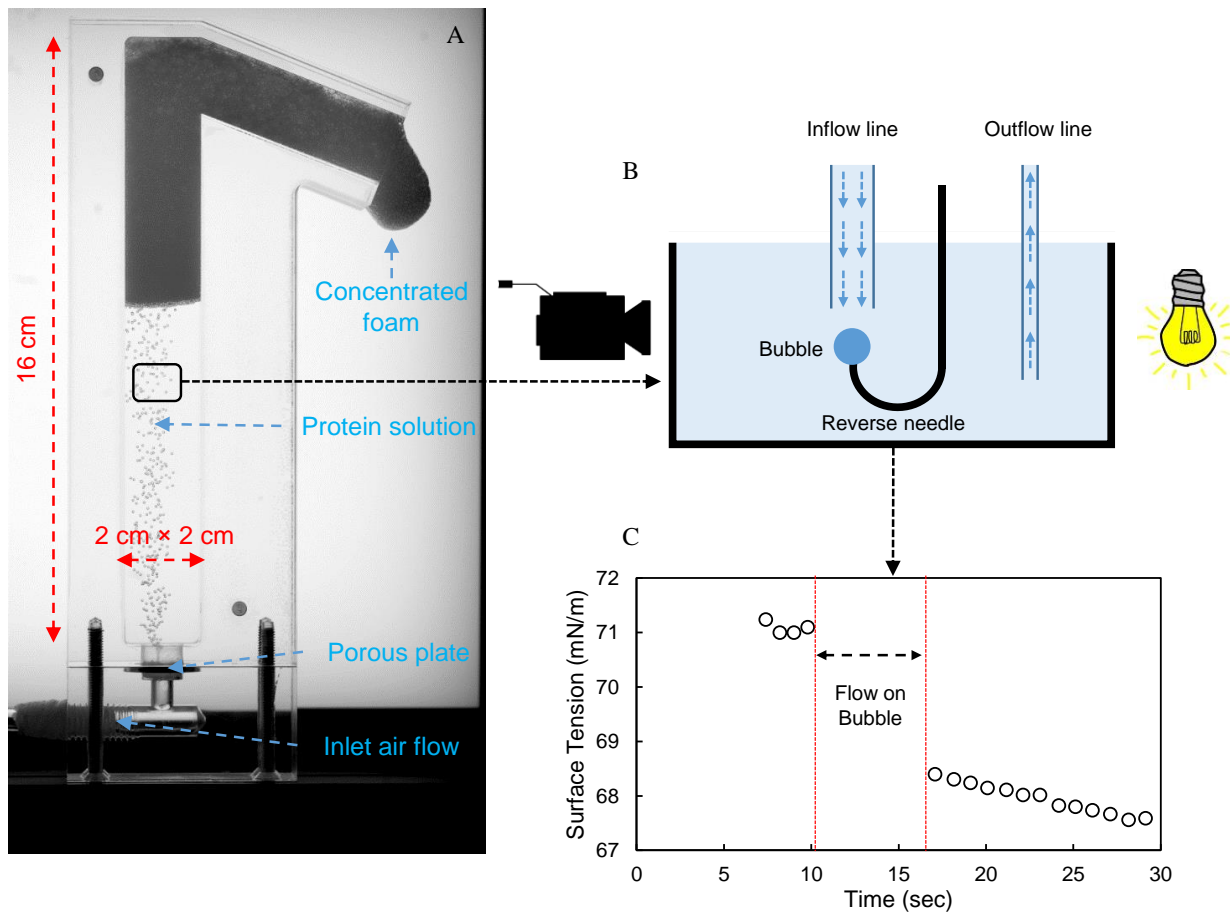

Figure 1. Top left: Schematic of the flow-on-bubble setup emulating the convective mass transfer at rising bubbles in the flotation column. The cuvette has an outer size of *35×35×32 mm (L×H×W)*. Bottom left: Surface tension before and after the flow on bubble. Right: Dimensions of the flotation cell used in this study.

## 2.4 Flotation experiments

The flotation experiments were performed in a Plexiglas cell (Figure 1-A) with an inner cross sectional area of 2 cm × 2 cm and a height of 16 cm. In each experiment, 10 cm of the column was filled with 40 ml of the BSA solution and the remaining 6 cm of the height is available to foam. The section where the foam flows over the weir has a height of 2 cm, so the foam exits the column at an average elevation of 15 ± 1 cm from the cell bottom (taken as efficient cell height). Bubbles are produced by injecting air through a sintered glass porous plate (pore size 40-100 µm, ROBU Glasfilter-Geraete GmbH, Germany) with a diameter of 20 mm and a thickness of 2.5 mm. The airflow rate was set using an Omega-500 mass flow controller (OMEGA Engineering GmbH, Germany). The high-speed camera Phantom VEO-410L (1280 x 720 pixel, High Speed Vision GmbH, Germany) was used to record the rising bubbles with a frame rate of 100 Hz. From the image sequence, an average rising velocity of the bubbles was obtained via the PIVLab plugin in MATLAB [19].

The porous glass plate was washed and dried before each run. Then, the cell was assembled and the air flow was initiated. Afterwards, the liquid was poured in the cell from the column top and the experiment was started. The extracted foam was collected at the outflow sequentially in tube samples and the foam weight was recorded with time. Concentration measurements were performed for the content of each of the tube samples.

## 2.5 Bicinchoninic acid (BCA) tests

The BCA assay is based on the formation of an intense purple complex consisting of $Cu^{1+}$ ions and bicinchoninic acid. $Cu^{1+}$ is formed by the reaction of proteins with alkaline $Cu^{2+}$ (Biuret reaction). The mentioned complex is stable and forms proportionally to the protein concentration [20]. In our experiments, the total BSA content was estimated using the Pierce™ BCA Protein Assay Kit (Thermo-Scientific, USA) following the manufacturer's protocol for the microplate procedure. For calibration, BSA solutions with concentrations of 0.025–1 mg/mL were used. For concentrations exceeding 1 mg/mL, the samples were diluted by buffer solution in order to keep the protein content in the measurement range.

# 3 Modeling description

## 3.1 Drainage equation

The following mass balance equation describes the relation between liquid flux and liquid content in an upward flowing foam in one dimension,

$$\frac{\partial \varphi}{\partial t} + \frac{\partial}{\partial z}\big(\varphi(u_z + u_g)\big) = 0 \qquad 1$$

where $\varphi$ is the liquid fraction, $t$ is time, $z$ is the foam height and $u_z$ is the $z$ component of the local liquid velocity in the plateau borders. Further, $u_g$ is the velocity of the foam bubbles which is related to the inlet gas volume flow rate ($Q_g$) and the cross sectional area of the flotation cell ($A_{col}$) as $u_g = \frac{Q_g}{A_{col}(1-\varphi)}$. The following equation describes the liquid flow through the foam [14, 21],

$$u_z = \frac{\alpha}{\eta \varphi}\left(-\rho g + \frac{\partial p_c}{\partial z}\right) \qquad 2$$

In the foam, a network of vertices and plateau borders acts like pores transferring the liquid [14]. In equation 2, $\eta$ is the fluid viscosity, $\rho$ is the liquid density, $g$ is the gravity acceleration, $\alpha$ is the permeability of the foam and $p_c$ is the pressure difference between air and water ($p_c = p_{air} - p_{water}$), i.e. the capillary pressure which is expressed in terms of surface tension $\gamma$ and plateau border radius of curvature $r$ as $p_c = \frac{\gamma}{r}$. The curvature of the plateau border is also related to the foam geometry through the following equation [14],

$$\frac{r}{R} = 1.73\varphi^{0.5} \qquad 3$$

where $R$ is the radius of a spherical bubble with the same volume as the bubble. Therefore, the capillary pressure is described as,

$$p_c = \frac{\gamma}{1.73R\varphi^{0.5}} \qquad 4$$

Combining equations 1 and 2, and applying the derivatives, yields

$$\frac{\partial \varphi}{\partial t} = \frac{\rho g}{\eta}\frac{\partial \alpha}{\partial z} - \frac{1}{\eta}\frac{\partial \alpha}{\partial z}\frac{\partial p_c}{\partial z} - \frac{\alpha}{\eta}\frac{\partial^2 p_c}{\partial z^2} - \frac{Q_g}{A_{col}(1-\varphi)^2}\frac{\partial \varphi}{\partial z} \qquad 5$$

Equation 5 is the drainage equation in the foam.

## 3.2 Solution method

A one dimensional computational domain is considered that treats the foam in a phase-averaged approach. The domain cross section is 2cm × 2cm and it has a height of 15 cm. At the start of the simulation, the domain contains 10 cm of liquid column corresponding to 40 ml of the main solution and 5 cm foam height with an initially homogeneous liquid content of 0.36. Equation 5 is solved numerically using a finite difference method. To that end, the domain is discretized with a node-centered grid into $n = 51$ grid cells. In every time step, the capillary pressure and permeability are calculated based on the liquid content values and the derivatives are estimated with central difference approximation. Then, we follow an Euler explicit time integration scheme to update the liquid content in each node at the next time step. A time step of 0.01 sec was used in the simulations. 0.001 sec time steps were also tried but did not change the simulation results more than 1%. A constant liquid content of 0.36 was set as boundary condition at the foam bottom, which corresponds to the liquid fraction of randomly packed spherical bubbles [14].

The liquid inflow from the foam top is zero, corresponding to free drainage. This can be implemented by setting equation 2 equal to zero, leading to

$$\rho g = \frac{\partial p_c}{\partial z} \qquad \qquad 6$$

Here, the term $\frac{\partial p_c}{\partial z}$ is calculated by applying the derivative on equation 4 as,

$$\frac{\partial p_c}{\partial z} = \frac{-\gamma}{1.73}\left(\frac{1}{R^2\,\varphi^{0.5}}\frac{\partial R}{\partial z} + \frac{0.5}{R\,\varphi^{1.5}}\frac{\partial \varphi}{\partial z}\right) \qquad \qquad 7$$

The liquid content derivative is estimated using backward differencing as $\frac{\partial \varphi}{\partial z} = \frac{\varphi_{top} - \varphi_{top-1}}{\Delta z}$. Therefore, by combining equations 6 and 7 and rearranging, the following equation is derived,

$$\frac{1.73\rho g}{\gamma} + \frac{1}{R^2 \varphi_{top}^{0.5}} \frac{\partial R}{\partial z} + \frac{0.5}{R \varphi_{top}^{1.5}} \frac{\varphi_{top} - \varphi_{top-1}}{\Delta z} = 0 \qquad 8$$

Equation 8 describes the formulation needed to calculate the liquid content at the top of the foam in the case of free drainage. Here, since the $\varphi$ values are already updated from bottom to top by solving equation 5, $\varphi_{top-1}$ is already known and $\varphi_{top}$ is the only unknown variable which is calculated using a Newton trial approach on equation 8. By determining $\varphi_{top}$, the rate of liquid extraction from the column is calculated as $Q_l = \frac{\varphi_{top}}{1-\varphi_{top}} Q_g$.

The initial height of the foam in the cell is 5 ± 1 cm. However, as the liquid is extracted from the cell, the foam-liquid contact point drops with time leading to a higher proportion of the foam column. This is implemented in the simulation by updating the liquid volume remaining in the cell at each time step and lowering the node point of the foam bottom.

### 3.3  Foam Permeability

The foam permeability depends on the liquid fraction, bubble radius and bubble surface mobility. Here, we use the Carman-Kozeny equation, which is described as follows,

$$\alpha = \frac{\varphi^3}{C_k A_s^2} \qquad 9$$

In this equation, $A_s$ is the specific interfacial area of the foam and $C_k$ is the Kozeny constant. Equation 9 considers a no-slip boundary condition at the bubble interface and is not applicable to mobile surfaces. However, for protein-laden interfaces, the assumption of zero velocity at the interface is reasonable [22] and satisfies the application of the Carman-Kozeny model. An estimation of $A_s$ for different liquid fractions is provided by Pitois et al. [15] who used the Surface Evolver software to calculate $A_s$ of a Kelvin foam at liquid fractions up to 0.32. With the obtained $A_s$ values, the permeability of the foam was estimated in [15] considering $C_k = 5$. Although the Kelvin foam does not exactly correspond to wet foams [14], the calculated permeabilities could still accurately predict the experimental results [15]. In this work, $A_s$ values were obtained from the plot reported by Pitois et al. [15].

## 4 Results

### 4.1 Flow-on-bubble results

Flow-on-bubble tests were performed at three different BSA concentrations of 0.01, 0.1 and 0.5 mg/mL corresponding to $1.5 \times 10^{-7}$, $1.5 \times 10^{-6}$ and $7.5 \times 10^{-6}$ M, respectively. The tests were performed after flow times of 2, 5, 10 and 20 seconds. The flow rates were 5, 10 and 20 ml/min corresponding to nozzle velocities of 1.7, 3.4 and 6.8 cm/sec, respectively. Higher nozzle velocities were not feasible, since otherwise the bubble would detach from the tip. The results are shown in Figure 2. Here the surface pressure (SP) is used which is the difference of the measured ST from the ST of pure water ($SP = 72.5 \, mN/m - ST$).

The SP values were extrapolated down to 1 sec to show the trend of the SP change at lower flow times. 1 sec is close to the rising time of the bubbles in our experiments. The slopes from the next closest data points were used for the extrapolation. The results show that the SP significantly increases with the flow velocity, as higher velocities considerably enhance the convective mass transfer around the bubble [22-26]. However, the influence is more prominent at lower concentrations. At constant flow velocity, the SP enhancement with time has a sharper rise at the beginning followed by a weaker increase. This implies that more adsorption occurs when the bubble surface is almost bare. However, the adsorption rate slows down as the stagnant cap covers the bubble [18, 24]. The sharp rise is well shown in the case of BSA 0.01 mg/mL while for higher concentrations, the rise occurs at very small times, which are not captured by our experiments. At constant flow time, the SP also increases with the flow velocity (Figure 3). The trend appears less pronounced at higher concentrations and the graph suggests that all curves are approaching the same limit around 12 ± 1 mN/m at larger flow velocities. This value might be considered as the maximum SP for the rising bubbles in BSA solution since we could not reach higher SPs even by applying the flow for a longer time. This value is yet smaller than the equilibrium SP = 17 ± 1 mN/m (according to figure S2 in the supplementary material (SM)), so reorientation processes or multi-layer adsorption might have a stronger influence in a quiescent long-time adsorption experiments.

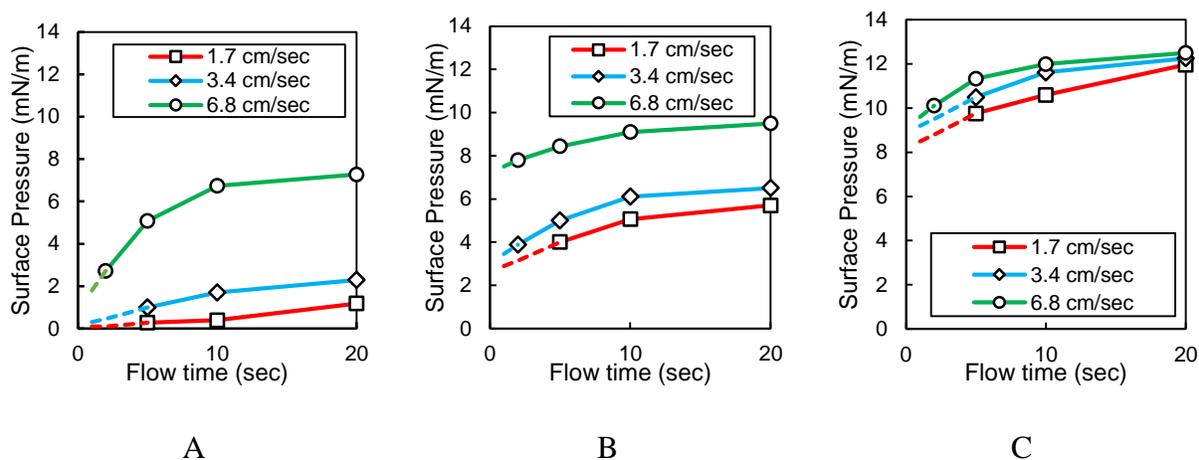

Figure 2. SP vs. flow time at different flow velocities and BSA concentrations of A) 0.01, B) 0.1 and C) 0.5 mg/mL.

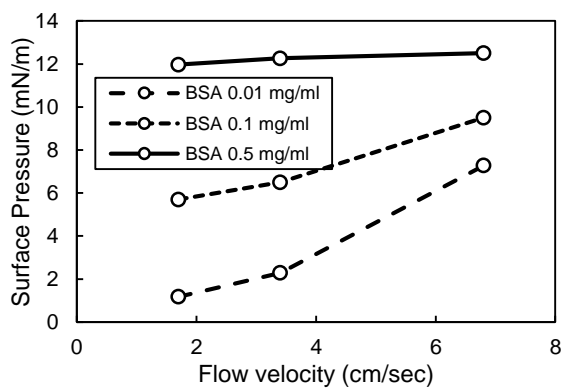

Figure 3. SP at flow time of 20 seconds vs. flow velocity for different BSA concentrations.

## 4.2 Flotation tests results

For each flotation experiment, the cell (Figure 1-A) was filled with 40 ml of the BSA solution and the foaming was initiated by applying 5 ml/min air flow. The tests were performed with BSA concentrations of 0.1, 0.25 and 0.5 mg/mL. When the air flow is started, the foam reaches the outlet after a lag time of 8-10 minutes. There, it is collected for liquid volume measurements and BCA tests. Figure 4-A shows the cumulative amount of recovered liquid and Figure 4-B gives the output concentration in different sampling time intervals. Average values from repeated runs are reported in these two figures. The liquid extract shows higher output volumes for higher BSA concentrations. This originates from the higher foam stability at higher BSA concentrations leading to wetter foams at the column top. The process here is mainly controlled by the foam coalescence, which occurs faster at lower concentrations. Considering in turn the enrichment at the outlet, the lowest BSA concentration shows the highest extract concentration. This is due to the fact that in the case of 0.1 mg/mL BSA, the foam is drier at the outlet while the foam surface concentration remains nearly constant. The final BSA bulk concentrations after approximately 90 min flotation time were $0.02 \pm 0.01$, $0.2 \pm 0.02$ and $0.42 \pm 0.04$ mg/mL, in comparison to the three initial bulk concentrations of 0.1, 0.25 and 0.5 mg/mL. The total amount of extracted BSA is also calculated and provided in SM-S9, which shows a higher amount of BSA extraction for the cases with more stable foam. However, the extraction efficiency is larger at lower initial concentrations.

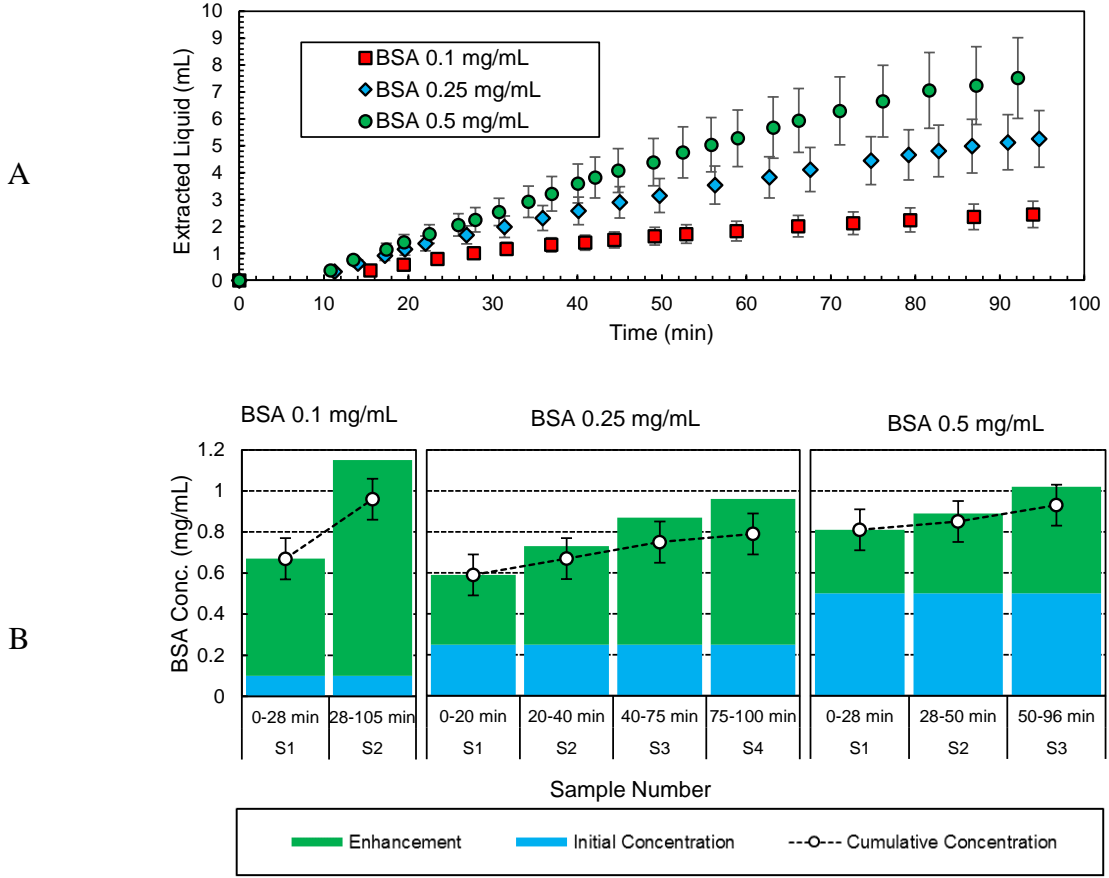

Figure 4. A) Total amount of recovered liquid in the flotation experiment for different BSA concentrations. B) BSA concentration in the foam extracted from the flotation cell measured by BCA method. Each sample belongs to a specified time span corresponding to the graphs in A. The cumulative concentration has been calculated by volume averaging of the consecutive samples. The concentrations have standard deviations smaller than ± 10%.

## 4.3 Flotation modeling results

In this section, different parameters required for the simulations are calculated and incorporated in the modeling. These parameters include the BSA adsorption on the rising bubble, BSA adsorption in the foam section and the bubble size in the foam column. With this, the simulations can be performed and the results are validated with the experimental tests. The BSA adsorption at the rising bubble is evaluated through flow-on-bubble experiments and the adsorption in the foam is estimated using dynamic ST tests. In both cases, a surface equation of state (EOS) is required to convert the ST data into BSA surface coverage [27]. The surface EOS defines the relationship between ST and surface coverage at the interface and is determined either theoretically or experimentally [28].

### 4.3.1 Modeling parameters estimation

#### 4.3.1.1 Surface EOS

The surface EOS was obtained in this study by performing compression tests on the bubble in the PAT device. The compression test takes advantage of the irreversible adsorption, considering a certain amount of protein which is already adsorbed at the interface. Then, the surface area is contracted to increase the surface coverage, while measuring the ST. This technique has previously been used in a Langmuir trough [28-29] by pouring a pre-determined amount of protein solution on the surface and applying the contractions. Here, we first formed an air bubble on the capillary tip of the PAT device and waited for a given time until a certain amount of BSA adsorbed on the bubble. Afterwards, the surface contractions were started by reducing the bubble size. A more detailed study on the dynamics and irreversibility of the BSA adsorption is provided in section 1 of SM. In addition, the details of the EOS calculation are provided in section 2 of the SM. The obtained EOS graph is plotted along with the standard deviation as error bars in Figure 5. This EOS significantly differs from the one calculated previously for BSA in Langmuir trough experiments by Hansen and Myrvold (1995) [28]. It is worth mentioning that in the Langmuir trough method, a predetermined amount of protein is spread on the interface assuming that no protein diffuses into the bulk. However, this assumption might lead to an overestimation of the surface coverage. In this work, we use our calculated EOS to determine the outlet concentration for the BSA flotation and draw a comparison between the two EOS approaches in the next sections.

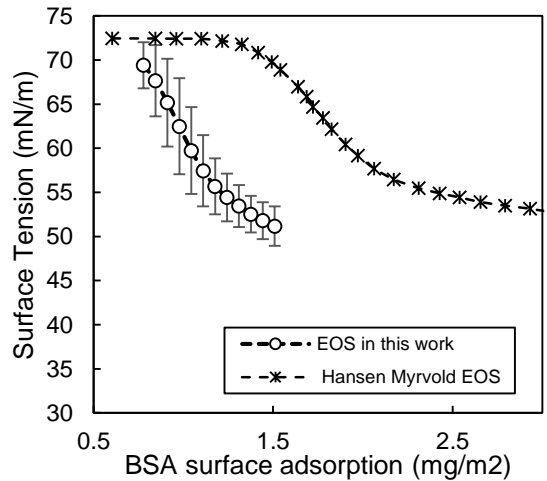

Figure 5. The calculated average EOS along with the EOS proposed by Hansen and Myrvold [28].

4.3.1.2 Adsorption on rising bubbles

The bubbles produced in the flotation cell were in the size range of 0.4-0.8 mm, as depicted in figure SM-S6. Their rising velocity was calculated by the PIVLab routine (available on MATLAB) [19]. Details on the determination of the rising velocity values are provided in figure SM-S7 for different experiments. The average rising velocity was calculated to $0.12 \pm 0.05$ m/s.

In this section, the flow-on-bubble results are used to estimate the ST of the rising bubbles in the flotation tests. Here the time and rising velocity of the bubbles as well as the BSA bulk concentration are the independent variables whose values determine the ST of the rising bubbles. The rising time is calculated by dividing the height of the remaining liquid in the cell by the average rising velocity. Finally, the rising velocity, rising time and the bulk BSA concentration are used to estimate the SP of the rising bubbles by linear interpolation and extrapolation of the graphs in Figure 2.

4.3.1.3 Adsorption in the foam section

After the rising bubble section, the surface adsorption continues in the foam section. There, we expect the ST to follow a trend as described in the dynamic ST experiments in Figure 6. From the tracking of individual foam bubbles, the residence time in the column is around 4 ± 1 minutes. With this, the final ST ($ST_2$) is calculated based on the initial ST ($ST_1$) and the bulk concentration as depicted in Figure 6 by the red arrows. The dynamic ST data for each bulk concentration is obtained by linear interpolation between the measured graphs.

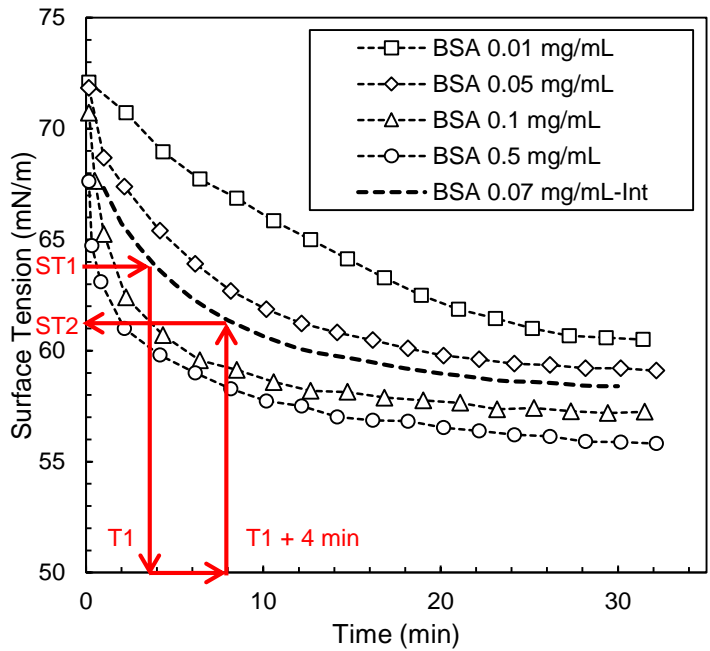

Figure 6. Dynamic ST of BSA solutions with concentrations of 0.01, 0.05, 0.1 and 0.5 mg/mL as well as an interpolated graph for BSA 0.07 mg/mL. The procedure for estimating the ST change in the foam is shown by red arrows.

4.3.1.4 Foam bubble size

The foam coarsening significantly influences the liquid drainage rate in flotation. Gas diffusion and coalescence are two important processes leading to bubble size growth in foams. Coalescence happens when the film separating two bubbles ruptures [30] while gas diffusion refers to the transfer of gas from the smaller bubbles with higher pressure to the larger ones with lower pressure leading to a gradual coarsening of the bubbles [14, 21, 31]. In this study, the residence time of the flowing foam in the flotation column is around 4 minutes. According to the high liquid content and relatively large bubble size of the foam, the gas diffusion is expected to occur at a much larger time scale [14, 32]. Therefore, the coalescence should determine the bubble size change in the foam column. However, we do not aim to focus on the mechanisms of bubble coarsening here, and only require a measure of the bubbles size as an input to improve our modeling. Therefore, we consider a linear bubble size growth with foam height $z$ as,

$$R(z) = R_0 + mz \qquad \qquad 10$$

where $R_0$ is the initial bubble size at the liquid-foam interface and $m$ is the slope of the bubble size change in the foam. The initial bubble size is obtained as an average value from the image analysis of the rising bubbles (figure SM-S6). In addition, the size change of the foam bubbles through the foam column is evaluated by analyzing the apparent surface film area size in contact with the column wall. In a foam, the actual bubble size slightly differs from the film size at the surface [32-33]. However, as long as only the size change is considered, the surface film analysis yields a suitable approximation of the actual bubble sizes. For that purpose, the picture of the foam column is firstly calibrated in ImageJ [34] by using the width of the column as a scale. Then, the foam column height is divided into 10 sections and an average bubble size is obtained in each section. An example is provided in figure SM-S8. Afterwards, a slope of the foam bubble size versus height is calculated as depicted in Figure 7. The foam images are evaluated at two different times during the experiment (early and late stage) and an average slope is calculated.

Table 1. Constants of equation 10

| *BSA concentration* | $R_0$ (mm) | m (m/m) |
|---|---|---|
| 0.1 mg/mL | $0.310 \pm 0.060$ | $0.019 \pm 0.003$ |
| 0.25 mg/mL | $0.295 \pm 0.045$ | $0.008 \pm 0.003$ |
| 0.5 mg/mL | $0.285 \pm 0.040$ | $0.005 \pm 0.000$ |

With the initial bubble size and the slope derived in Figure 7, a linear relation for $R(z)$ is obtained for every concentration which is input in the modeling. Figure 7 shows the slowest foam bubble growth for the case of 0.5 mg/mL BSA, and the fastest growth for the case of 0.1 mg/mL BSA. The reason for that is the decrease in concentration during the experiment from 0.1 to around 0.02 mg/mL, as reported earlier. This reveals that the bulk is almost free of BSA molecules at the end. The time dependence is not resolved in the simulations, but is included in the uncertainty range of the slope (see Table 1).

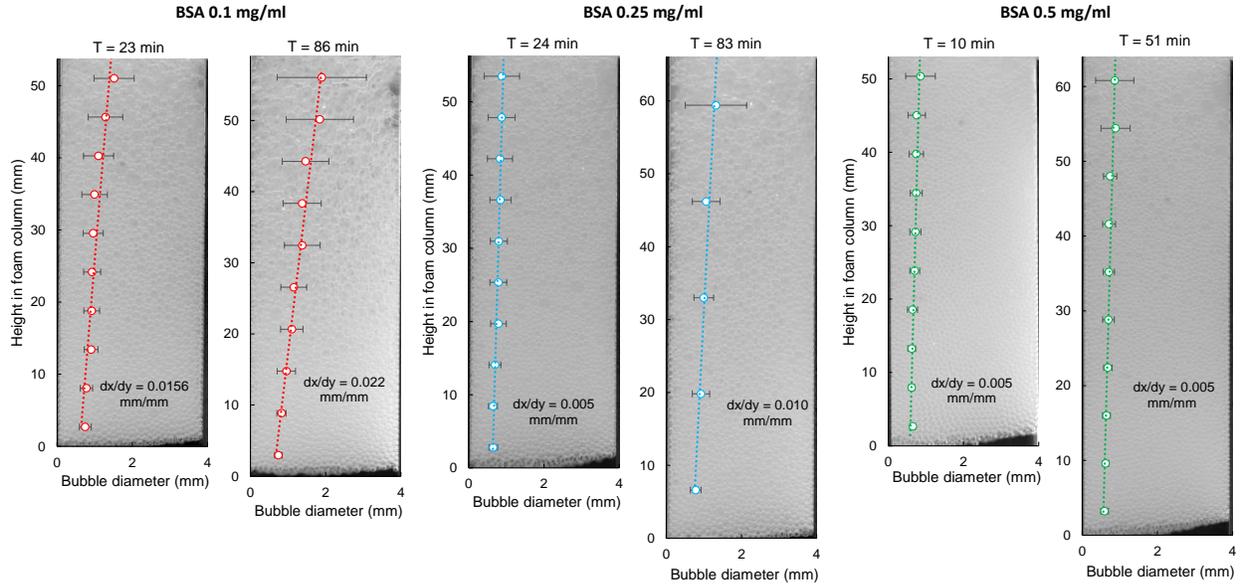

Figure 7. Height dependence of bubble size in the foam column for different BSA concentrations at different times. The error bars quantify the standard deviation of the bubble size at every height.

### 4.3.2 Modeling results

The simulations were run based on the procedure given in section 3.2 using the parameters summarized in Table 2. According to the results presented in Figure 8, the volume (top row) and the BSA concentration (bottom row) of the extracted liquid are predicted well by the simulation results within a reasonable error range. The average values of the parameters of foam bubble size (Table 1), surface EOS (Figure 5) and the column height (Table 2) were input in the simulation. In addition, the given uncertainties were used for an overall error estimation in the simulations, yielding the dashed lines in Figure 8 as the upper and lower uncertainty levels.

Table 2. Simulation parameters.

| Simulation parameters | Value |
|---|---|
| Water viscosity ($\eta$) | 0.001 Pa.s |
| Water density ($\rho$) | 1000 kg/m3 |
| Total column length | 15 ± 1 cm |
| Initial foam column length | 5 ± 1 cm |
| Column cross sectional area ($A_{col}$) | 4 cm2 |
| BSA molecular weight | 66400 g/mol |
| Initial liquid volume | 40 ml |
| Air flow rate ($Q_g$) | 5 ml/min |
| Initial BSA concentration | 0.1, 0.25 and 0.5 mg/mL |
| Time step ($\Delta t$) | 0.01 sec |

According to the results, the simulation tends to underestimate the BSA concentration at the outlet. This is more prominent for the cases of 0.1 and 0.25 mg/mL BSA. Here, one reason could be the overestimation of the foam liquid content. Indeed, for the case of 0.5 mg/mL BSA initial concentration, where the coalescence is less, a better agreement between simulation and experimental results is observed.

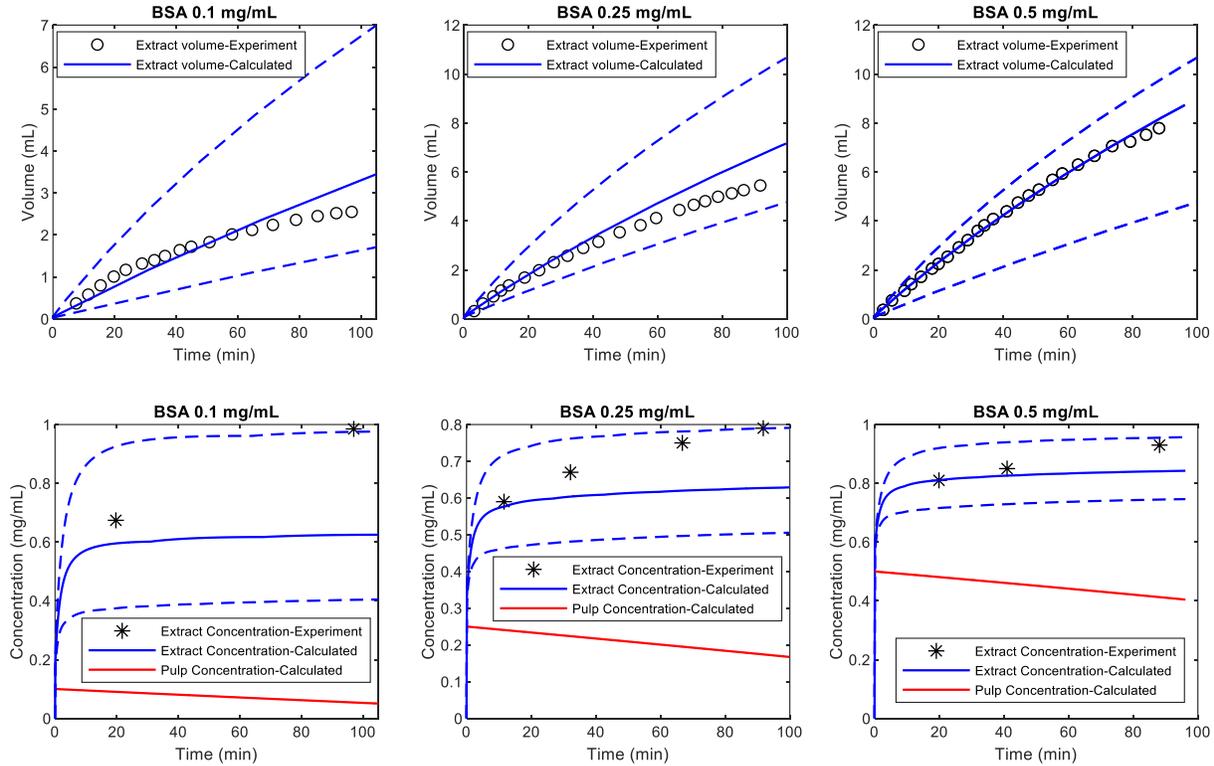

Figure 8. Simulation and experimental results; Top row) extracted liquid volume and bottom row) BSA concentration in the extracted liquid for initial BSA concentrations of 0.1, 0.25 and 0.5 mg/ml, respectively. The dashed lines describe the upper and lower uncertainty levels as estimated in the simulation.

The SP values over time for the three concentrations are reported in figure SM-S10. The calculated SP results from the gradual SP build up in the rising bubble section and further adsorption in the foam. The simulations of the cases with 0.25 and 0.5 mg/mL initial BSA concentration do not show a considerable change in SP with time while for 0.1 mg/mL, a significant reduction is observed due to the major bulk depletion. This is also reflected in the notable reduction in foam stability for the case of 0.1 mg/mL BSA. The profiles of liquid fraction, capillary pressure and permeability at the end of the simulations are provided in figures S11-S13 of the SM.

## 5   Discussion

This work shows successful BSA enrichment in the flotation process. Here, the BSA protein itself provides the required foamability and foam stability and eliminates the need of surfactant addition. The solution pH of 7.3, which is close to BSA neutral point, intensifies the proteins affinity to the interface [35]. Hence, foam flotation provides a non-contaminating biocompatible approach to concentrate the protein from aqueous solutions. Foam stability directly affects the recovery and grade. A lower foam stability produces a drier foam and leads to a higher grade of the extract. In this regard, the adsorption process plays a crucial role since almost the same final surface coverage is achieved even at lower bulk concentrations. The almost constant final surface coverage (described in figure SM-S2) originates from the irreversible adsorption of BSA at the interface. Nevertheless, for an optimized flotation enrichment, the foam should be stable enough to flow over the weir and to provide a sufficient foam height for drainage of the liquid. Therefore, an optimization of the foam stability is required.

In addition, an adequate modeling was presented in this work offering an efficient and cost effective tool to simulate the protein flotation process. In the next step, it allows us to optimize the flotation operational parameters such as gas flow rate, column height and bubble size. Moreover, once the modeling is validated with laboratory flotation tests, it can be scaled up to larger volumes. Our approach in this work was to deliver the necessary input parameters of the modeling for a satisfactory prediction of the flotation recovery and grade. To that end, a linear dependence of the foam coarsening on the height was measured during the experiment and was incorporated in the modeling. The obtained slope is an average value over the whole period and tends to overestimate the flotation recovery at later times of experiments. Therefore, a refined approach could incorporate a time dependent foam-coarsening which takes into account the impact of the concentration reduction in the bulk solution.

The protein grade in the extract, on the other hand, depends not only on the liquid content but also on the protein surface coverage of the foam. In the flotation column, the protein adsorption occurs both in the bulk solution and the foam. The rising time of the bubble in the bulk solution is in the order of 1 second. In this part, the bubble rises with a high velocity, depending on its size, until it reaches the solution top and joins the foam. In the rising bubble part, the flow field around the bubble induces a convective mass transfer of the protein to the bubble surface. The adsorbed material is then swept by the flow to the bottom part of the bubble, and a stagnant cap with immobile interface forms. This immobile part induces a higher drag force on the bubble resulting in a reduced bubble velocity [18, 22-25, 17, 28]. The bubble surface is fully immobilized by further adsorption, so that the rising velocity corresponds to Stokes' law [18, 22]. Since the amount of adsorbed surfactant on rising bubbles cannot be directly characterized in the rising bubble experiments, many works have instead evaluated the adsorption by the change in rising bubble velocity [17, 18, 22]. However, this method is not applicable to an immobile interface since the rising velocity becomes independent of the adsorption after the stagnant cap fully covers the bubble surface. The adsorption also continues in the stagnant cap. This effect is particularly important for small bubbles, which get immobile soon [18]. Hence, in this work, we have utilized the flow-on-bubble technique to estimate the surface coverage regardless of the surface mobility. The flow-on-bubble tests emulate the convective mass transfer and provide the surface coverage by measuring the surface tension of the bubble after the flow is applied. The bubble afterwards resides in the foam section for 4-5 minutes. Here, the adsorption takes place under significantly reduced convection, which is emulated using the dynamic surface tension measurement results. The modeling predicts the average outlet concentration well. However, the temporal concentration evolution is not captured by the simulations. This is because the foam is assumed to have a constant foam stability during the simulation, while in reality the foam gets coarser with time, resulting in a dryer extract and a higher BSA concentration at the outlet. Furthermore, the film breakage in the foam could also influence the surface coverage in the foam section since the adsorbed material on a rupturing lamella could be transferred to the bulk solution and/or in the neighboring lamellae. However, this is not yet well understood and further investigation is required to capture the mass transfer after a lamella rupture, which will be the subject of future work.

In addition to the mentioned points, for a successful adsorption estimation, an EOS is necessary to translate the ST into adsorption data. The EOS was obtained by performing surface compression tests in PAT where an irreversible adsorption of BSA molecules was assumed. Here, we compare the grade of the flotation extract predicted by the modeling using the calculated EOS in this work to the EOS reported by Hansen and Myrvold [28]. The case of 0.5 mg/mL BSA, where the numerical flotation recovery agrees well with the experiments (cf. Figure 8-right column), was chosen for validation to exclude the influence of recovery on the grade. According to Figure 9, the Hansen and Myrvold EOS gives a higher outlet concentration, while the results corresponding to our EOS match the experiments better. As explained previously in section 4.3.1.1, a predetermined amount of protein is spread on the interface in the Langmuir trough method used in [28], assuming no protein diffusion to the bulk. However, this assumption might lead to an overestimation of adsorption, contributing to the deviation in Figure 9.

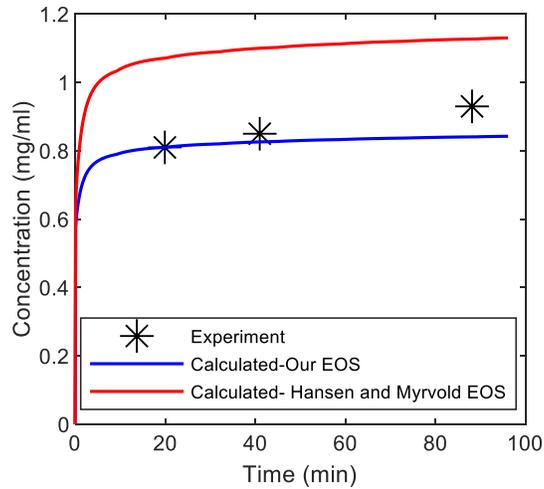

Figure 9. Outlet BSA concentration vs. time at 0.5 mg/mL initial BSA concentration for two different surface EOSs. Both EOSs are depicted in Figure 5.

# 6   Conclusions

Former research works on the application of flotation for protein separation have been mainly focused on the experimental aspects of the operational parameters [3-9] while modeling approaches are mostly lacking. An attempt for the modeling was made by Du et al (2000) who calculated the upward and downward liquid flow in the foam using approximations based on the local liquid hold up which was directly measured in the experiments [10]. In the present work, on the other hand, we described a more comprehensive modeling of the protein flotation where the drainage equation was solved using basic approximations of the capillary pressure and liquid permeability in the foam, without requiring a prior knowledge on the liquid hold up. A novelty of this work is that no tuning parameter are employed in the modeling, but all ingredients are derived from adequately designed pre-experiments. Measuring bubble size in the liquid column and in the foam yields the initial surface area and a quantity for the foam coarsening. An estimation of the protein adsorption on rising bubbles is provided through flow-on-bubble tests. This is required to calculate the protein concentration in the foam, i.e. in the extracted foam. The flow-on-bubble tests further revealed that the surface coverage never reaches its equilibrium value even at a high concentration of 0.5 mg/ml. This is in contrast with the assumption of equilibrium at the bubble surface assumed in previous works [10].

A key finding of the work was the relationship between the BSA outlet concentration and the foam stability. With lower foam stability, the rapid foam coarsening led to an accelerated liquid drainage and a dryer foam at the outlet. However, in batch flotation this might be difficult to achieve, because the ongoing depletion of the bulk solution might change the foam stability significantly [32].

The work will be continued with the separation of enzymes using flotation where the time scale of the separation process plays an important role according to the possible denaturation and activity loss of the enzyme. In addition, the modeling will be upgraded with the addition of a more advanced foam coarsening model.

**Declaration of Competing Interest**

The authors declare that they have no known competing financial interests or personal relationships that could have appeared to influence the work reported in this paper.

**Acknowledgements:**

This work was supported by the Deutsche Forschungsgemeinschaft (DFG) with the grant numbers HE 7529/2-1, HE 7529/3-1 and AN 387/7. We also acknowledge financial support provided by the Helmholtz Zentrum Dresden-Rossendorf (HZDR).

**References**

1. Shean, B.J. and Cilliers, J.J., 2011. A review of froth flotation control. International Journal of

Mineral Processing, 100(3-4), pp.57-71.

2. Rubio, J., Souza, M.L. and Smith, R.W., 2002. Overview of flotation as a wastewater treatment technique. Minerals engineering, 15(3), pp.139-155.

3. Schügerl, K., 2000. Recovery of proteins and microorganisms from cultivation media by foam flotation. In New Products and New Areas of Bioprocess Engineering (pp. 191-233). Springer, Berlin, Heidelberg.

4. Bhattacharjee, S., Kumar, R. and Gandhi, K.S., 1997. Prediction of separation factor in foam separation of proteins. Chemical engineering science, 52(24), pp.4625-4636.

5. Gehle, R.D. and Schügerl, K., 1984. Protein recovery by continuous flotation. Applied microbiology and biotechnology, 20(2), pp.133-138.

6. Mukhopadhyay, G., Khanam, J. and Nanda, A., 2010. Protein removal from whey waste by foam fractionation in a batch process. Separation science and technology, 45(9), pp.1331-1339.

7. Li, R., Wu, Z., Wangb, Y., Ding, L. and Wang, Y., 2016. Role of pH-induced structural change in protein aggregation in foam fractionation of bovine serum albumin. Biotechnology Reports, 9, pp.46-52.

8. Lockwood, C.E., Bummer, P.M. and Jay, M., 1997. Purification of proteins using foam fractionation. Pharmaceutical Research, 14(11), pp.1511-1515.

9. Linke, D., Zorn, H., Gerken, B., Parlar, H. and Berger, R.G., 2007. Laccase isolation by foam fractionation—new prospects of an old process. Enzyme and Microbial Technology, 40(2), pp.273-277.

10. Du, L., Loha, V. and Tanner, R.D., 2000. Modeling a protein foam fractionation process. In Twenty-First Symposium on Biotechnology for Fuels and Chemicals (pp. 1087-1099). Humana Press, Totowa, NJ.

11. Otzen, D., 2011. Protein–surfactant interactions: a tale of many states. Biochimica et Biophysica Acta (BBA)-Proteins and Proteomics, 1814(5), pp.562-591.

12. Saha, D., Ray, D., Kohlbrecher, J. and Aswal, V.K., 2018. Unfolding and refolding of protein by a combination of ionic and nonionic surfactants. ACS omega, 3(7), pp.8260-8270.

13. Lotfi, M., Bastani, D., Ulaganathan, V., Miller, R. and Javadi, A., 2014. Bubble in flow field: A new experimental protocol for investigating dynamic adsorption layers by using capillary pressure tensiometry. Colloids and Surfaces A: Physicochemical and Engineering Aspects, 460, pp.369-376.

14. Cantat, I., Cohen-Addad, S., Elias, F., Graner, F., Höhler, R., Pitois, O., Rouyer, F. and Saint-Jalmes, A., 2013. Foams: structure and dynamics. OUP Oxford.

15. Pitois, O., Lorenceau, E., Louvet, N. and Rouyer, F., 2009. Specific surface area model for foam permeability. Langmuir, 25(1), pp.97-100.

16. Karbaschi, M., Bastani, D., Javadi, A., Kovalchuk, V.I., Kovalchuk, N.M., Makievski, A.V., Bonaccurso, E. and Miller, R., 2012. Drop profile analysis tensiometry under highly dynamic conditions. Colloids and Surfaces A: Physicochemical and Engineering Aspects, 413, pp.292-297.

17. Dukhin, S.S., Kovalchuk, V.I., Gochev, G.G., Lotfi, M., Krzan, M., Malysa, K. and Miller, R., 2015. Dynamics of Rear Stagnant Cap formation at the surface of spherical bubbles rising in surfactant solutions at large Reynolds numbers under conditions of small Marangoni number and slow sorption kinetics. Advances in Colloid and Interface Science, 222, pp.260-274.

18. Zhang, Y., McLaughlin, J.B. and Finch, J.A., 2001. Bubble velocity profile and model of surfactant mass transfer to bubble surface. Chemical engineering science, 56(23), pp.6605-6616.

19. Thielicke, W. and Stamhuis, E., 2014. PIVlab–towards user-friendly, affordable and accurate digital particle image velocimetry in MATLAB. Journal of open research software, 2(1).

20. Smith, P.E., Krohn, R.I., Hermanson, G.T., Mallia, A.K., Gartner, F.H., Provenzano, M., Fujimoto, E.K., Goeke, N.M., Olson, B.J. and Klenk, D.C., 1985. Measurement of protein using bicinchoninic acid. Analytical biochemistry, 150(1), pp.76-85.

21. Stevenson, P. ed., 2012. Foam engineering: fundamentals and applications. John Wiley & Sons.

22. Ybert, C. and Di Meglio, J.M., 1998. Ascending air bubbles in protein solutions. The European Physical Journal B-Condensed Matter and Complex Systems, 4(3), pp.313-319.

23. Fleckenstein, S. and Bothe, D., 2013. Simplified modeling of the influence of surfactants on the rise of bubbles in VOF-simulations. Chemical engineering science, 102, pp.514-523.

24. Zholkovskij, E.K., Koval'chuk, V.I., Dukhin, S.S. and Miller, R., 2000. Dynamics of rear stagnant cap formation at low Reynolds numbers: 1. Slow sorption kinetics. Journal of colloid and interface science, 226(1), pp.51-59.

25. Ybert, C. and Di Meglio, J.M., 2000. Ascending air bubbles in solutions of surface-active molecules: influence of desorption kinetics. The European Physical Journal E, 3(2), pp.143-148.

26. Eftekhari, M., Schwarzenberger, K., Heitkam, S. and Eckert, K., 2021. Interfacial flow of a surfactant-laden interface under asymmetric shear flow. Journal of Colloid and Interface Science, 599, pp.837-848.

27. Rosen, M.J. and Kunjappu, J.T., 2012. Surfactants and interfacial phenomena. John Wiley & Sons.

28. Hansen, F.K. and Myrvold, R., 1995. The kinetics of albumin adsorption to the air/water interface measured by automatic axisymmetric drop shape analysis. Journal of Colloid and Interface Science, 176(2), pp.408-417.

29. Cho, D., Narsimhan, G. and Franses, E.I., 1996. Adsorption dynamics of native and alkylated

derivatives of bovine serum albumin at air–water interfaces. Journal of colloid and interface science, 178(1), pp.348-357.

30. Neethling, S.J. and Cilliers, J.J., 2003. Modelling flotation froths. International Journal of Mineral Processing, 72(1-4), pp.267-287.

31. Hilgenfeldt, S., Koehler, S.A. and Stone, H.A., 2001. Dynamics of coarsening foams: accelerated and self-limiting drainage. Physical review letters, 86(20), p.4704.

32. Hoang, D.H., Heitkam, S., Kupka, N., Hassanzadeh, A., Peuker, U.A. and Rudolph, M., 2019. Froth properties and entrainment in lab-scale flotation: A case of carbonaceous sedimentary phosphate ore. Chemical Engineering Research and Design, 142, pp.100-110.

33. Wang, Y. and Neethling, S.J., 2009. The relationship between the surface and internal structure of dry foam. Colloids and Surfaces A: Physicochemical and Engineering Aspects, 339(1-3), pp.73-81.

34. Hartig, S.M., 2013. Basic image analysis and manipulation in ImageJ. Current protocols in molecular biology, 102(1), pp.14-15.

35. Yamasaki, M., Yano, H. and Aoki, K., 1990. Differential scanning calorimetric studies on bovine serum albumin: I. Effects of pH and ionic strength. International journal of biological macromolecules, 12(4), pp.263-268.